\documentclass{PoS}

\newcommand{\asls}{\mbox{$a_{\rm sl}^{s}$}}
\def\plotdir{figures/}

\title{LHCb Semileptonic Asymmetry}

\ShortTitle{LHCb Semileptonic Asymmetry}

\author{\speaker{Mika Vesterinen}\thanks{On behalf of the LHCb Collaboration.}\\ 
        CERN\\
        E-mail: \email{mika.vesterinen@cern.ch}}

\abstract{
  A recent LHCb measurement of the $CP$ violating flavour specific asymmetry in $B_s^0$ decays, \asls, is presented.
  This measurement is based on a data sample corresponding to an integrated luminosity of 1~fb$^{-1}$ of $pp$ collisions at $\sqrt{s}$ $=$ $7$~TeV, 
  collected during the 2011 run of the LHC.
  The result is $\asls = (-0.24 \pm 0.54_{\rm stat} \pm 0.33_{\rm syst})\times 10^{-2}$, which is the most precise measurement
  of this quantity to date and agrees with the Standard Model prediction.}

\FullConference{14th International Conference on B-Physics at Hadron Machines,\\
		April 8-12, 2013\\
		Bologna, Italy }

\begin{document}

\section{Introduction}

Precision measurements of flavour observables offer a possible window to new physics,
that is complementary to direct searches for on-shell production of new particles.
Thanks to the large $b\bar{b}$ cross section at the LHC, we are entering a
new regime in sensitivity with these studies.
The $B_s^0$ sector in particular is not accessible by $e^+e^-$ colliders
operating at the $\Upsilon(4S)$ resonance.
Since the flavour eigenstates,  $B_s^0$ and $\bar{B}_s^0$ of this meson
are not eigenstates of the weak Hamiltonian,
an initial $B_s^0$ state will evolve into a mixture of $B_s^0$ and $\bar{B}_s^0$.
The following $2 \times 2$ complex matrix characterises this so called ``mixing'':
\begin{equation}
\left( \begin{array}{cc} M_{11} - \frac{i}{2}\Gamma_{11} & M_{12} - \frac{i}{2}\Gamma_{12} \\[0.5em] M_{12}^* - \frac{i}{2}\Gamma_{12}^* & M_{22} - \frac{i}{2}\Gamma_{22}  \end{array} \right).
\end{equation}
The mass and decay width differences between the mass eigenstates are denoted
$\Delta M_s$ and $\Delta\Gamma_s$, respectively.
Violation of $CP$ in mixing would be apparent as a flavour specific asymmetry, e.g., in semileptonic decays,
\newcommand{\GammaWSBbar}{\Gamma(\bar{B^0_s} \rightarrow \mu^+ D_s^-)}
\newcommand{\GammaWSB}{\Gamma(B_s^0 \rightarrow \mu^- D_s^+)}
\begin{equation}
a_{\rm sl}^s \equiv \frac{\GammaWSBbar - \GammaWSB}{\GammaWSBbar + \GammaWSB} = \frac{\Delta\Gamma}{\Delta M}\tan\phi_{12},
\label{Equation:asl}
\end{equation}
where $\phi_{12}=\arg(-M_{12}/\Gamma_{12})$.
The Standard Model predicts a very small value for $\phi_{12}$ -- around $0.2^{\circ}$~\cite{LenzNierste2007}.
Therefore, the predicted value of $a_{\rm sl}^s = (1.9 \pm 0.3)\times 10^{-5}$ is negligible compared to current experimental precision.
A measurement that is significantly different from zero would be a strong indication of new physics.

The D0 Collaboration reported an anomalously large asymmetry in the rate of like-sign muon pairs,
$A_{\mu\mu} = (-0.787 \pm 0.172_{\rm stat} \pm 0.093_{\rm syst})\times 10^{-2}$~\cite{DzeroAmumu}.
This observable is a linear combination of the semileptonic asymmetries of the $B_s^0$ and $B_d^0$ systems:
$A_{\mu\mu} \approx 0.6a_{\rm sl}^s + 0.4a_{\rm sl}^d$.
They have also recently measured the separate asymmetries~\cite{DzeroAsls,DzeroAsld}, and the results are consistent with the Standard Model.
The most precise measurement of $a_{\rm sl}^{d}$ from the BaBar Collaboration~\cite{BaBarAsld}
is also consistent with the Standard Model.

The measurement of $a_{\rm sl}^s$ with equation~\ref{Equation:asl} requires flavour tagging to 
select wrong-sign decays.
Without flavour tagging, the time integrated asymmetry is,
\begin{equation}
A_{\rm meas} \equiv   \frac{N(\mu^+ D_s^-) - N(\mu^- D_s^+)}{N(\mu^+ D_s^-) + N(\mu^- D_s^+)} = \frac{a_{\rm sl}^s}{2} + \mathcal{K}\left[a_{p} - \frac{a_{\rm sl}^s}{2} \right],
\label{Equation:ASL_UNTAGGED}
\end{equation}
where $a_p$ is the production asymmetry, and 
\begin{equation}
\mathcal{K} = \frac{\int e^{-\Gamma_s t}\cos(\Delta M_st)\epsilon(t)dt}{\int e^{-\Gamma_s t}\cosh(\Delta \Gamma_s t/2)\epsilon(t)dt}.
\end{equation}
The production asymmetry arises due to the asymmetric ($pp$) initial state and the
forward geometry of the LHCb detector, and has been measured to be around 1\%
for the $B_s^0$~\cite{BsAP}.
Inserting the LHCb efficiency as a function of decay time, $\epsilon(t)$, and the mixing parameters for the $B_s^0$,
one obtains a value of $\mathcal{K} \approx 2\times 10^{-3}$.
A production asymmetry of even a few percent is thus washed out by the fast oscillations, to a level that is negligible
compared to the statistical precision on $A_{\rm meas}$.
The decay mode $B_s^0 \rightarrow D_s^- \mu^+\nu_{\mu}X$ is used, with the $D_s^-$ decaying to $\phi\pi^-$ and $\phi \rightarrow K^+K^-$.
In order to obtain $A_{\rm meas}$, the signal yields need to be corrected for any instrumental asymmetries.

\section{\label{Sec:Data}Dataset and event selection}

This analysis~\cite{ASL_CONF} is based on 1~fb$^{-1}$ of $pp$ collisions at $\sqrt{s} = 7$~TeV, 
collected with the LHCb detector~\cite{LHCbDetector} during the 2011 run of the LHC.
LHCb is a single-arm forward spectrometer covering the pseudorapidity
range $2 < \eta < 5$, designed for the study of particles containing $b$ or $c$ quarks. 
The magnet polarity is reversed periodically. Approximately 40\% of the data were
recorded with the magnetic field pointing up and the rest with the magnetic field pointing down. We exploit the fact
that certain detection asymmetries cancel if data from different magnet polarities are
combined. 
The trigger consists of a hardware stage (L0), based on
information from the calorimeter and muon systems, followed by a software stage (HLT) which
applies a full event reconstruction.

The offline selection requires a three track vertex, consistent with the decay $D_s^+ \rightarrow K^+K^-\pi^+$.
The $D_s^+$ candidate must make a vertex with a muon that is consistent with the decay of a $B_s^0$ meson.
All of the daughter particles must have a significant impact parameter with respect to the nearest primary vertex.
The kaons must satisfy particle identification requirements and the kaon pair must have an invariant
mass within $\pm$20~MeV of the $\phi$(1020) resonance.
Signal candidates are triggered by the muon at L0 and at the first stage of the HLT.
The L0 muon trigger performs a fast reconstruction of muons, 
and requires a transverse momentum, $p_T$, of at least 1.4 GeV (the exact threshold varied during the data taking period).
The first HLT stage, HLT1, requires this muon to be confirmed in the tracking stations and to have
a significant impact parameter with respect to the nearest primary vertex.
In the final HLT stage, HLT2, an inclusive, topological trigger for $b$ hadron decays with a muon 
in the final state is used~\cite{Hlt2BTopo}.

\section{\label{Sec:Backgrounds}Backgrounds}

\begin{figure}\centering
\includegraphics[width=0.8\linewidth]{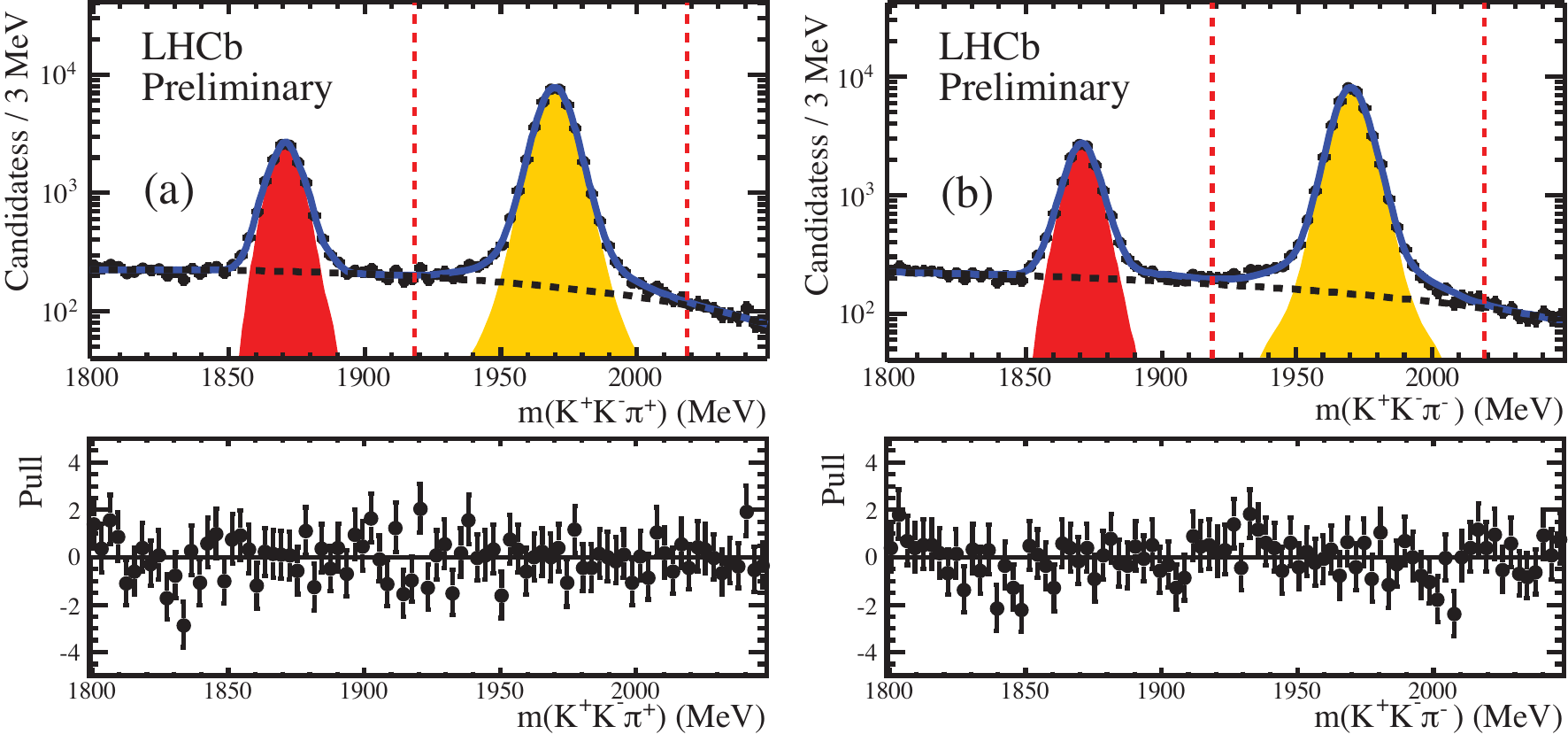}
\caption{The $K^+K^-\pi^{\pm}$ invariant mass distributions of the (a) $D_s^+\mu^-$ and (b) $D_s^-\mu^+$ candidates in the 
magnet-up dataset.}
\label{Fig:asl_mass}
\end{figure}

Random combinatoric background is subtracted by fitting the $K^+K^-\pi^{\pm}$ invariant mass distribution.
Figure~\ref{Fig:asl_mass} shows a fit to roughly half of the dataset, that was collected with the magnet polarity in the 
up configuration.
The background from prompt $D_s^+$ production is estimated to be around 3\% from a two-dimensional fit
to the mass and logarithm of the impact parameter of the $D_s^+$.
The $D_s^+$ production asymmetry has been measured to be $(0.33 \pm 0.22_{\rm stat} \pm 0.11_{\rm syst})\times 10^{-2}$~\cite{DsAP}.
The effect of prompt $D_s^+$ on the measured asymmetry is therefore considered to be negligible.
A contamination of less than 1\% is estimated from $\bar{B}^0_s \rightarrow D_s^+ X$, with a hadron faking a muon.
The fake rates are measured to have small asymmetries, so this background has a negligible effect
on the measurement.
The decay of $B^0/B^{\pm} \rightarrow D_s^{\pm} X_c$, where the second charm hadron, $X_c$, decays semileptonically to produce a muon
is estimated to contribute at the few percent level.
The process $B^- \rightarrow D_s^+ K^- \mu^- \bar{\nu} X$ also contributes at the few percent level.
Both of these sources inherit possible $B^0$ and $B^{\pm}$ production asymmetries of a percent at the most,
but with opposite sign.
A small systematic uncertainty is assigned to cover the residual asymmetry.

\section{\label{Sec:ADet}Instrumental asymmetries}

Any detection asymmetry is mostly cancelled by averaging the two magnet polarity 
configurations.
Rather than to rely on this cancellation, dedicated control channels are used to 
measure the detection asymmetries, and separately correct the signal yields 
in each polarity.

Muon identification and trigger efficiency asymmetries are measured using
$J/\psi \rightarrow \mu^+\mu^-$ candidates.
Both muons are reconstructed in the tracking detectors, such the the full decay
kinematics are determined, but only one (tag) muon is required to be
identified in the muon stations. 
One can then study the efficiencies using the unbiased (probe) muon.

\begin{figure}\centering
\includegraphics[width=0.55\linewidth]{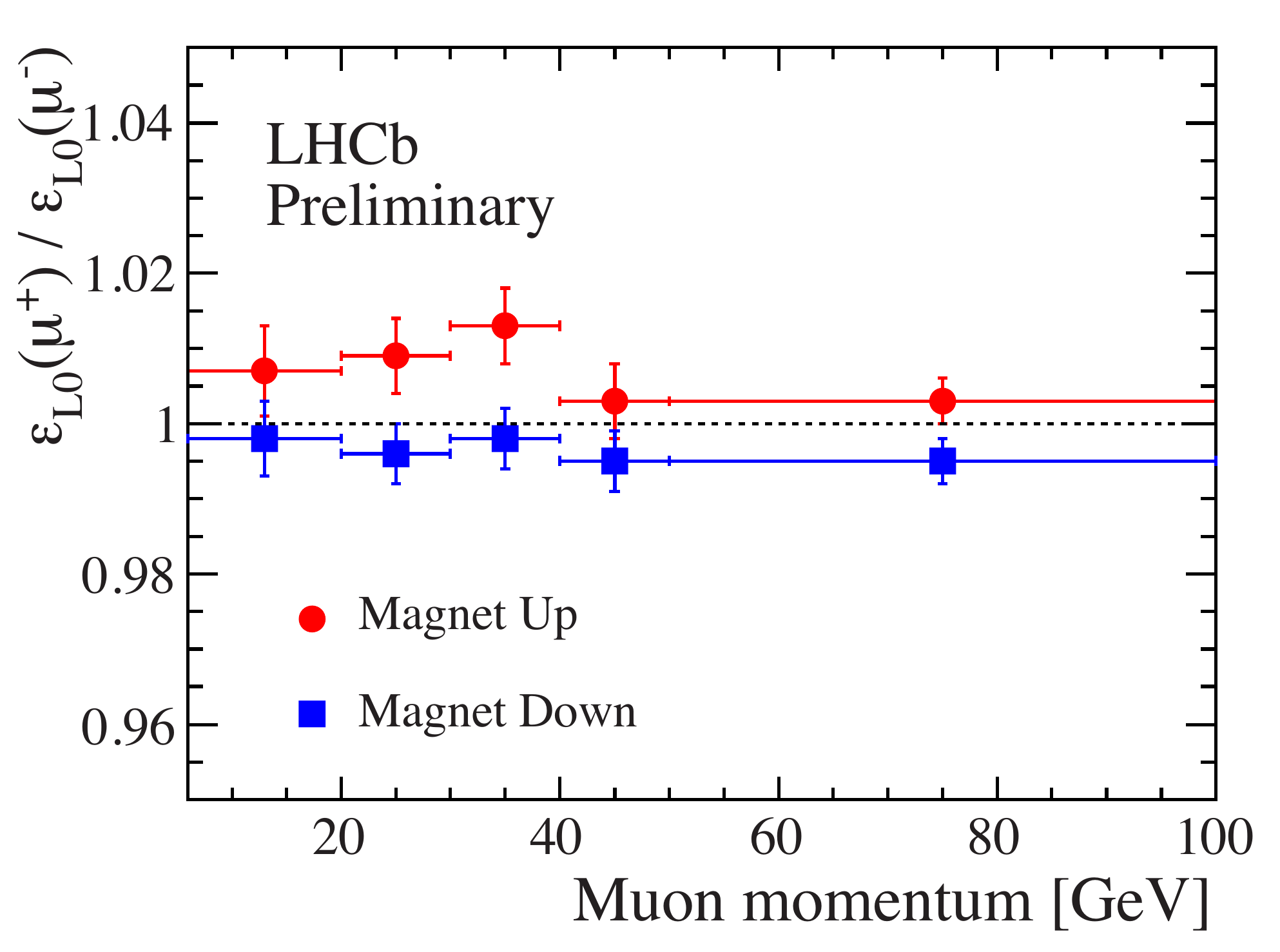}
\caption{The L0 muon trigger efficiency ratio ($\mu^+/\mu^-$) as a function of the offline 
reconstructed muon momentum. The red and blue points correspond to the magnet-up and magnet-down
data sets, respectively.}
\label{Fig:L0asy}
\end{figure}

Figure~\ref{Fig:L0asy} shows the L0 muon trigger efficiency ratio ($\mu^+/\mu^-$) as a function of the offline 
reconstructed muon momentum. 
For a single magnet polarity, an asymmetry of order 1\% is observed, but is reasonably well compensated
by the other polarity, as expected.
In order to account for any kinematic dependence of the muon asymmetries, 
the measurement is performed in a 50 bin grid of momentum, $p_x$ and $p_y$.
The $J/\psi$ based muon corrections account for the offline muon identification,
and the first two stages of the trigger.


A partially reconstructed sample of $D^{*+} \rightarrow D^{0}\pi^+$ decays, with $D^0 \rightarrow K^- \pi^+ \pi^+ \pi^-$,
is used to determine the $\pi^+$ detection asymmetry,
as detailed in Ref~\cite{DsAP}.
Since the muon and the pion have opposite charge, their tracking asymmetries are mostly cancelled.
A residual asymmetry may remain due to their different average momenta,
and is accounted for in the systematic uncertainties.
A possible asymmetry in HLT2 is studied using a sample of $\bar{B}^0 \rightarrow D^+ \mu^- \bar{\nu} X$ 
candidates, with $D^+ \rightarrow K^- \pi^+ \pi^+$.
No significant asymmetry is found, and the upper limit is propagated to the systematic uncertainty.

\section{\label{Sec:Systematics}Systematic uncertainties}

Table~\ref{Table:syst} lists the sources of systematic uncertainty 
on $A_{\rm meas}$. 
The largest component is the statistical uncertainty on the muon efficiency corrections,
due to the limited $J/\psi$ sample size.
The next largest sources are from varying the signal fit parameterisation,
and from the residual tracking asymmetries due to the different kinematics of the 
muon and the pion.

\begin{table}
\centering
\caption{\label{Table:syst}Absolute systematic uncertainties in $A_{\rm meas}$.}
\begin{tabular}{lc}
\hline\hline
Source & $\sigma(A_{\rm meas})$ (\%) \\
\hline
Signal modelling & 0.06 \\
Background from other $b$ hadrons & 0.05\\
Tracking asymmetries & 0.06\\
Kaon asymmetries & 0.02\\
Muon corrections & 0.05\\
Varying run conditions & 0.01\\
Muon mis-identification & 0.01\\
HLT2 biases & 0.05 \\
Statistical uncertainty in the muon corrections & 0.10\\
\hline
Total & 0.16\\
\hline\hline
\end{tabular}
\end{table}

\section{\label{Sec:Results}Results}

\begin{figure}
\centering
\includegraphics[width=\linewidth]{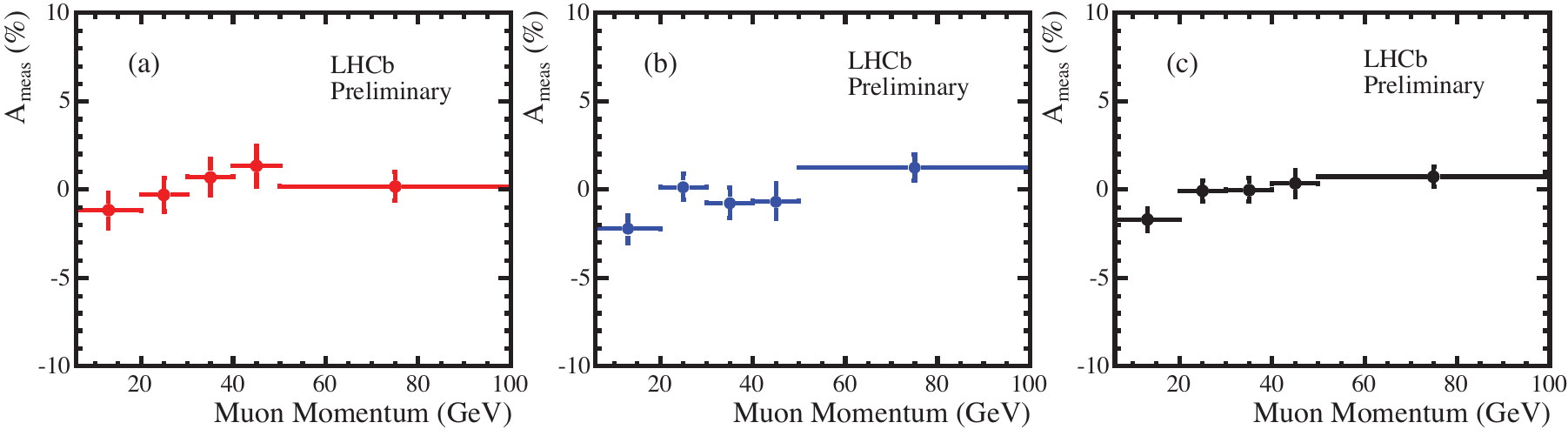}
\caption{$A_{\rm meas}$ as a function of muon momentum for (a) magnet up, (b) magnet down, and (c) the average.}
\label{Fig:Ameas}
\end{figure}

\begin{figure}
\centering
\includegraphics[width=0.55\linewidth]{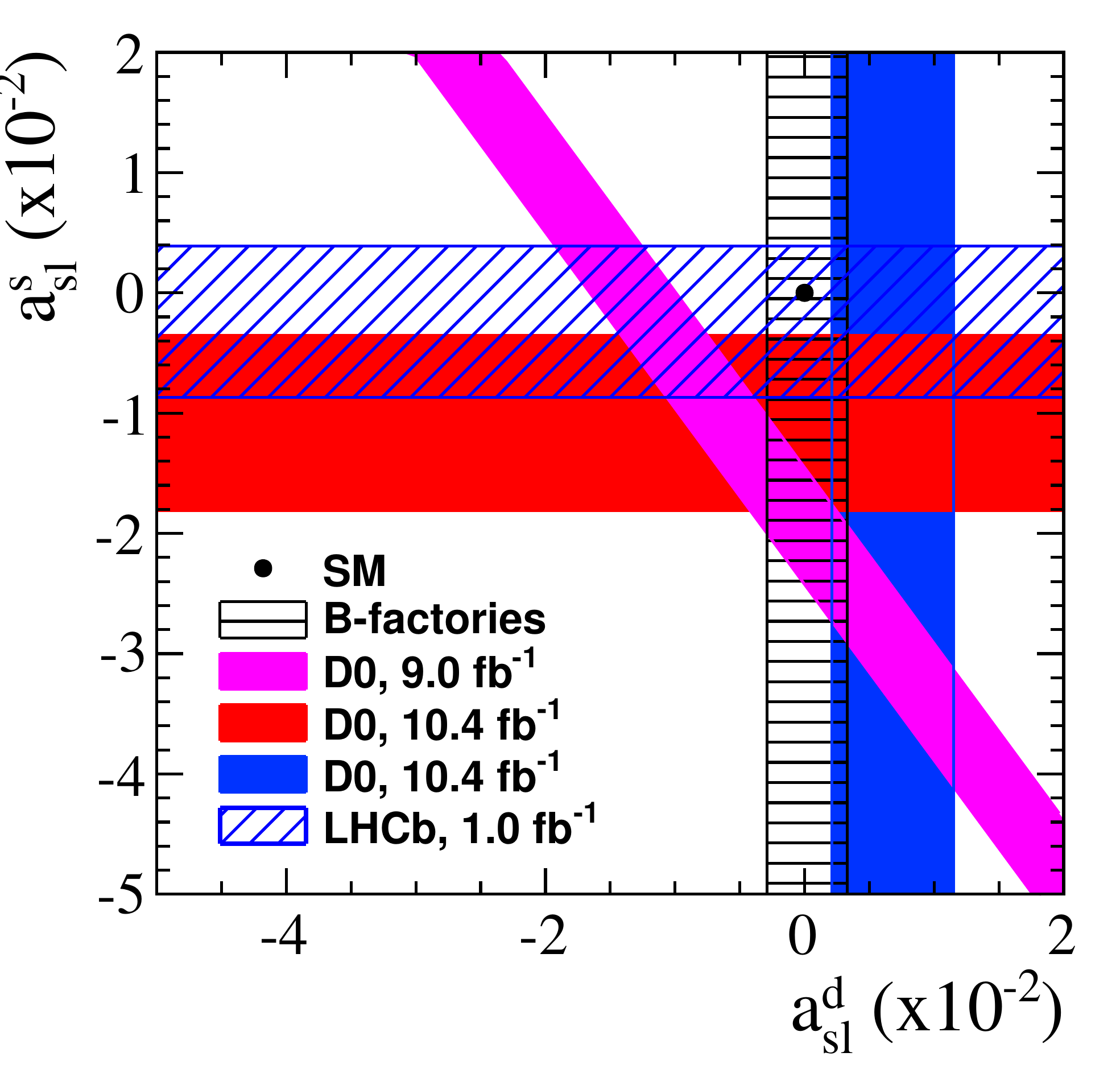}
\caption{Comparison of the different measurements of $a_{\rm sl}^s$ and $a_{\rm sl}^d$.}
\label{Fig:asl_comparison}       
\end{figure}

Figure~\ref{Fig:Ameas} shows $A_{\rm meas}$ as a function of the muon momentum,
separately for the two magnet polarities and also for their average.
Averaging over momentum we measure,
\[ a_{\rm sl}^s = (-0.24 \pm 0.54_{\rm stat} \pm 0.33_{\rm syst})\times 10 ^{-2}.\] 
Figure~\ref{Fig:asl_comparison} compares this result to other measurements of 
$a_{\rm sl}^s$ and $a_{\rm sl}^d$, plus the D0 measurement of $A_{\mu\mu}$.
This is the most precise measurement to date and is in good agreement with the Standard Model prediction,
thus not confirming the anomaly seen by the D0 Collaboration.

\section{Conclusions}

A recent measurement of the flavour specific $CP$ violating asymmetry in $B_s^0$ decays, $a_{\rm sl}^s$  is presented.
This measurement is based on the full 2011 dataset of 1~fb$^{-1}$ of $pp$ collisions at $\sqrt{s}=7$~TeV.
The result, $a_{\rm sl}^s = (-0.24 \pm 0.54_{\rm stat} \pm 0.33_{\rm syst}) \times 10^{-2}$, is
the most precise measurement of this quantity to date, and is good agreement with Standard Model expectations.
This result does not confirm the anomalous dimuon asymmetry reported by the D0 Collaboration.

\end{document}